\documentclass[floatfix, nopacs, twocolumn, showkeys, preprintnumbers, nofootinbib, superscriptaddress]{revtex4-1}
\usepackage[utf8]{inputenc}
\usepackage[sort&compress]{natbib}
\usepackage{ulem}
\usepackage{overpic}
\usepackage{bm}
\usepackage{times}
\usepackage{amssymb,amsbsy,amsmath,amsfonts}
\usepackage{graphicx}
\usepackage{color}
\usepackage{booktabs}
\usepackage{makecell}

\usepackage{rotating}
\usepackage{srcltx}
\usepackage{slashed}
\usepackage{subfigure}
\usepackage{multirow}
\usepackage{verbatim}

\usepackage{tabularx}

\usepackage{hyperref}

\DeclareUnicodeCharacter{3002}{HEREHEREHERE}

\begin{document}

\title{Triaxial shape of the one-proton emitter $^{149}$Lu}

\author{Qi Lu}
\affiliation{School of Physics, Beihang University, Beijing 102206, China}

\author{Kai-Yuan Zhang}\email{zhangky@caep.cn}
\affiliation{Institute of Nuclear Physics and Chemistry, China Academy of Engineering Physics, Mianyang, Sichuan 621900, China}

\author{Shi-Sheng Zhang}\email{zss76@buaa.edu.cn}
\affiliation{School of Physics, Beihang University, Beijing 102206, China}

\date{\today}

\begin{abstract}
We revisit the proton emitter $^{149}$Lu utilizing the recently developed triaxial relativistic Hartree-Bogoliubov theory in continuum (TRHBc).
By incorporating the microscopic nuclear structure properties from the TRHBc theory into the WKB approximation, we successfully reproduce the measured proton-emission half-life of $^{149}$Lu within experimental uncertainties.
A triaxial ground state characterized by ($\beta=0.17,\gamma=31^\circ$) has been clarified for $^{149}$Lu.
The inclusion of triaxiality significantly changes nuclear density distributions and potentials, which results in enhanced binding of both the nuclear system and the proton-emitting orbital.
As a result, a slightly extended half-life for the proton emission of $^{149}$Lu is achieved after considering triaxial deformation degrees of freedom.
\end{abstract}


\maketitle

\section{Introduction}
\label{sec:Introduction}  

Exotic nuclei near the dripline might exhibit novel features such as nuclear halos~\cite{PhysRevLett.55.2676, PhysRevLett.112.242501}, changes of magic numbers~\cite{PhysRevLett.84.5493}, and emissions of nucleons~\cite{PhysRevLett.108.102501, PhysRevLett.132.082501}, etc.
These arouse great attention from nuclear structure and inevitable impact on nucleosynthesis in nuclear astrophysics~\cite{HeM2020AJ,XuSZ2021EPJA}.

Proton radioactivity is one of the most appealing phenomena observed in nuclei beyond the proton dripline~\cite{DELION2006113}.
So far, more than 30 proton emitters have been identified from experiments~\cite{Woods:1997cs,BLANK2008403,Pfutzner:2011ju, PhysRevLett.128.112501}. 
Among those, the one-proton emitter $^{149}$Lu was discovered lately with the shortest directly measured half-life of $T_{1/2} = 450^{+170}_{-100}$ nanoseconds (ns)~\cite{PhysRevLett.128.112501}. 
Such a short and accurate half-life puts stringent constraints on the test of various nuclear theories.

Actually, the description of proton radioactivity is rather challenging due to its sensitive dependence on many factors, including decay energies, potential barriers, and spectroscopic factors~\cite{DELION2006113}.
Numerous efforts have been devoted to evaluate the half-lives of proton emitters~\cite{PhysRevC.61.067301,PhysRevC.69.054311,PhysRevC.64.034317,PhysRevC.58.3011,PhysRevLett.81.538,PhysRevLett.79.2217,PhysRevLett.82.4595}.
Recently, based on the proton-nucleus potential and the spectroscopic factors from the deformed relativistic Hartree-Bogoliubov theory in continuum (DRHBc)~\cite{Zhou:2009sp, Li:2012gv} that assumes axially symmetric shape for atomic nuclei, the half-life of $^{149}$Lu was reproduced within experimental uncertainties by using the WKB approximation~\cite{XIAO2023138160}.
This success is attributed to the fact that the DRHBc theory incorporates deformation, pairing correlations~\cite{MengX2020PRC}, and continuum effects in a microscopic and self-consistent way, which are essential to describe exotic nuclei near the dripline~\cite{Meng:2015hta,Sun:2018ekv,Zhang:2019qeu,Sun:2020tas,Zhang2020PRC, Sun:2021nyl,Sun:2021alk,Yang:2021pbl,Zhang2021PRC(L),Pan2021PRC,He2021CPC,Zhong2022-uj,Zhang2022ADNDT,Pan2022PRC,Zhang:2023dhj,Zhang:2023bqg,An2024PLB,Guo2024ADNDT}.

For $^{149}$Lu, the potential coexistence of prolate and oblate deformations has been indicated by both the nonadiabatic quasiparticle model~\cite{PhysRevLett.128.112501} and the DRHBc theory~\cite{XIAO2023138160}.
However, its possible triaxial deformation has not been taken into account yet previously.
We notice that many proton-rich Lu isotopes including $^{161}$Lu~\cite{Bringel2005EvidenceFW}, $^{163}$Lu~\cite{PhysRevLett.86.5866,PhysRevLett.89.142503}, $^{165}$Lu~\cite{SCHONWAER20039}, and $^{167}$Lu~\cite{AMRO2003197} have been reported to manifest the wobbling mode, which is a fingerprint of triaxiality~\cite{Bohr1975NuclearSV}.
Therefore, we are triggered to check if $^{149}$Lu is triaxially deformed and evaluate the possible impact on the proton-emission half-life with newly developed triaxially deformed relativistic Hartree-Bogoliubov theory in continuum (TRHBc)~\cite{Zhang2023PRCL2}, which inherits all advantages of the DRHBc theory and further includes the triaxial deformation degrees of freedom.

In this Letter, we briefly introduce the theoretical framework in Sect. \ref{sec:Frame}, discuss the results in Sect. \ref{sec:Results},
and finally make a summary in Sect. \ref{sec:summary}.

\section{Theoretical Framework}
\label{sec:Frame} 
\subsection{The WKB approximation}

Following Refs.~\cite{2022EPJWC.26011037X,XIAO2023138160}, we employ the WKB approximation to calculate the proton emission half-life,
\begin{align}\label{eq:halflife}
    T_{1/2} = \frac{\hbar \ln{2}}{S_F \it{\Gamma}}, 
\end{align}
where $S_F$ is the spectroscopic factor and $\it{\Gamma}$ is the decay width.
For a parent nucleus with its valence proton occupying a certain orbital, $S_F$ can be estimated as~\cite{Sorensen:1966zz}
\begin{align}\label{eq:Sfactor}
    S_F =u^2,
\end{align}
where $u^2$ denotes the nonoccupation probability of this orbital in the daughter nucleus and can be obtained self-consistently in the TRHBc framework. 
The decay width $\it{\Gamma}$ is calculated by averaging ${\it{\Gamma}}(\theta,\phi)$ in all directions,
\begin{equation}\label{eq:finalWidth}
{\it{\Gamma}} = \frac{\int {\it{\Gamma}}(\theta,\phi) d\Omega}{ 4\pi},
\end{equation}
with $\theta$ being the orientation angle and $\phi$ being the azimuthal angle of the emitted proton with respect to the daughter nucleus~\cite{Qian:2016teg}. 

In the WKB approximation, ${\it{\Gamma}}(\theta,\phi)$ is calculated by
\begin{align}\label{eq:width}
    {\it{\Gamma}}(\theta,\phi) = \mathcal{N} \frac{\hbar^2}{4\mu}\exp{\left(-2\int_{r_2}^{r_3}k(r,\theta,\phi) dr\right)},
\end{align}
in which $\mathcal{N}$ refers to the quasi-classical bound-state normalization factor,
$\mu$ denotes the reduced mass of the emitted proton and the daughter nucleus,
and $k(r,\theta,\phi)$ is related to the decay energy $E$ and potential barrier $V(r,\theta,\phi)$ via
\begin{equation}
k(r,\theta,\phi) = \sqrt{2\mu|E - V(r,\theta,\phi)|} .
\end{equation}
Here, $r_i$ represents the $i$-th turning point of the potential barrier with respect to the decay energy $E$.
The normalization factor $\mathcal{N}$ is defined as~\cite{Buck:1992zza}
\begin{align}\label{eq:nor}
    \frac{1}{\mathcal{N}(\theta,\phi)} = \int_{r_1}^{r_2} \frac{dr}{k(r,\theta,\phi)} \cos^2 \left( \int_{r_1}^r dr' k(r',\theta,\phi)  -\frac{\pi}{4} \right).
\end{align}

The potential barrier includes the nuclear, Coulomb, and centrifugal contributions,
\begin{align}\label{eq:barrier}
    V(r,\theta,\phi) = V_N(r,\theta,\phi) + V_C(r,\theta,\phi) + \frac{\hbar^2}{2\mu} \frac{l(l+1)}{r^2},
\end{align}
where $V_{N/C}(r,\theta,\phi)$ refers to the nuclear/Coulomb potential and $l$ denotes the orbital angular momentum of the emitted proton, also self-consistently obtained in TRHBc calculations.

\subsection{The TRHBc theory}
The potential barrier is crucial for the calculation of proton-emission half-life.
As the TRHBc theory is capable of describing triaxial exotic nuclei near the drip line \cite{Zhang2023PRCL2}, it is adopted to provide nuclear and Coulomb potentials.
In the TRHBc theory, the relativistic Hartree-Bogoliubov (RHB) equation reads~\cite{peter1991}

\begin{align}\label{eq:RHB}
    \left(\begin{array}{cc}
        h_D - \lambda & \Delta \\
        -\Delta^* & -h_D^* + \lambda 
    \end{array}\right)
    \left(\begin{array}{c}
        U_k \\
        V_k
    \end{array}\right)
    = E_k \left(\begin{array}{c}
        U_k \\
        V_k
    \end{array}\right),
\end{align}
in which $h_D$ is the Dirac Hamiltonian, $\lambda$ denotes the Fermi surface, $\Delta$ represents the pairing potential, and $E_k$ and $\left(U_k,V_k\right)^T$ respectively refer to the quasiparticle energy and wave function.
In coordinate space,
\begin{align}\label{eq:hd}
    h_D (\bm{r})= \bm{\alpha}\cdot\bm{p} + V(\bm{r}) + \beta \left[M + S(\bm{r})\right],
\end{align}
with the vector potential $V(\bm{r})$ and the scalar potential $S(\bm{r})$ constructed by various densities, and
\begin{align}\label{eq:pairing}
    \Delta(\bm{r_1},\bm{r_2}) = V^{pp}(\bm{r_1},\bm{r_2})\kappa (\bm{r_1},\bm{r_2}),
\end{align}
with $V^{pp}$ and $\kappa$ being the pairing interaction and the pairing tensor, respectively.
Following the DRHBc theory, the present TRHBc theory employs the zero-range density-dependent pairing interaction,
\begin{align}
    V^{pp} (\bm{r_1},\bm{r_2}) = V_0 \frac{1-P^{\sigma}}{2} \delta(\bm{r_1}-\bm{r_2})\left( 1-\frac{\rho(\bm{r_1})}{\rho_{\text{sat}}}\right),
\end{align}
where $V_0$ is the pairing strength, $\rho_{\text{sat}}$ represents the saturation density of nuclear matter, and the operator $(1-P^\sigma)/2$ projects onto the spin-zero component. 

To incorporate triaxial deformation degrees of freedom, potentials and densities are expanded in terms of spherical harmonics,
\begin{align}
    f(\bm{r}) = \sum_{lm} f_{lm}(r)Y_{lm}(\Omega).
\end{align}
Due to the assumed spatial reflection symmetry and mirror symmetries respect to $xy$, $xz$, and $yz$ planes, the expansion orders $l$ and $m$ are restricted to be even numbers, and the radial components satisfy the condition $f_{lm}(r) = f_{l-m}(r)$ \cite{Xiang2023Symmetry}.

Nuclei near the drip line may exhibit significantly diffuse density distributions.
To describe properly such possible large spatial extensions, the RHB equation \eqref{eq:RHB} should be solved in a basis that has appropriate asymptotic behavior at large $r$.
To this end, a Dirac Woods-Saxon basis \cite{Zhou:2003jv,Zhang2022PRC}, which turns out to be successful in many nuclear models~\cite{Long2010PRC,Zhou2010PRC(R),Geng2020PRC,Geng2022PRC}, is adopted in the TRHBc theory.

\section{Results and discussion}
\label{sec:Results}
In the previous work \cite{XIAO2023138160}, several relativistic density functionals were employed in the DRHBc calculations, and the results were found almost density-functional independent.
Therefore, in this work we take the density functional NL3$^*$ \cite{Lalazissis:2009zz} as an example, and other numerical details in our calculations remain the same as those employed in Ref. \cite{XIAO2023138160}.

\begin{figure}[htpb]
\includegraphics[width=0.45\textwidth]{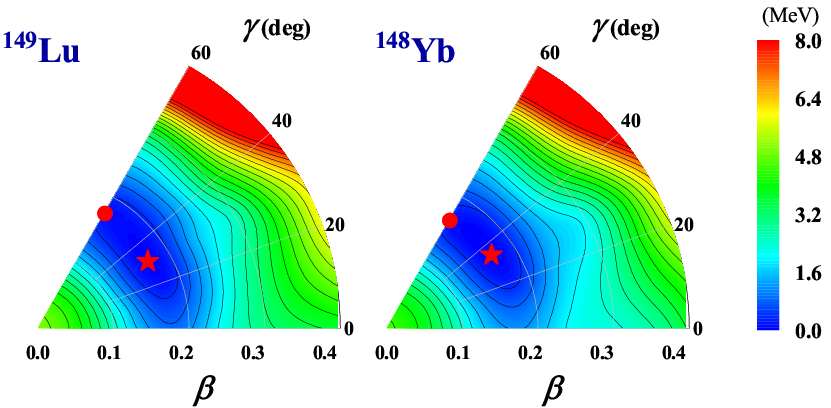}
\caption{Potential energy surfaces for $^{149}$Lu and $^{148}$Yb in the $\beta$-$\gamma$ plane from the TRHBc theory. The energy separation between each contour line is 0.4 MeV. All energies are normalized with respect to the energy of absolute minimum (triaxial ground state) indicated by a red star. The ground-state deformation predicted by the DRHBc theory is denoted by a red closed circle.}
\label{fig:potentialenergy}
\end{figure}

\begin{figure*}[htpb]  
  \includegraphics[width=0.8\textwidth]{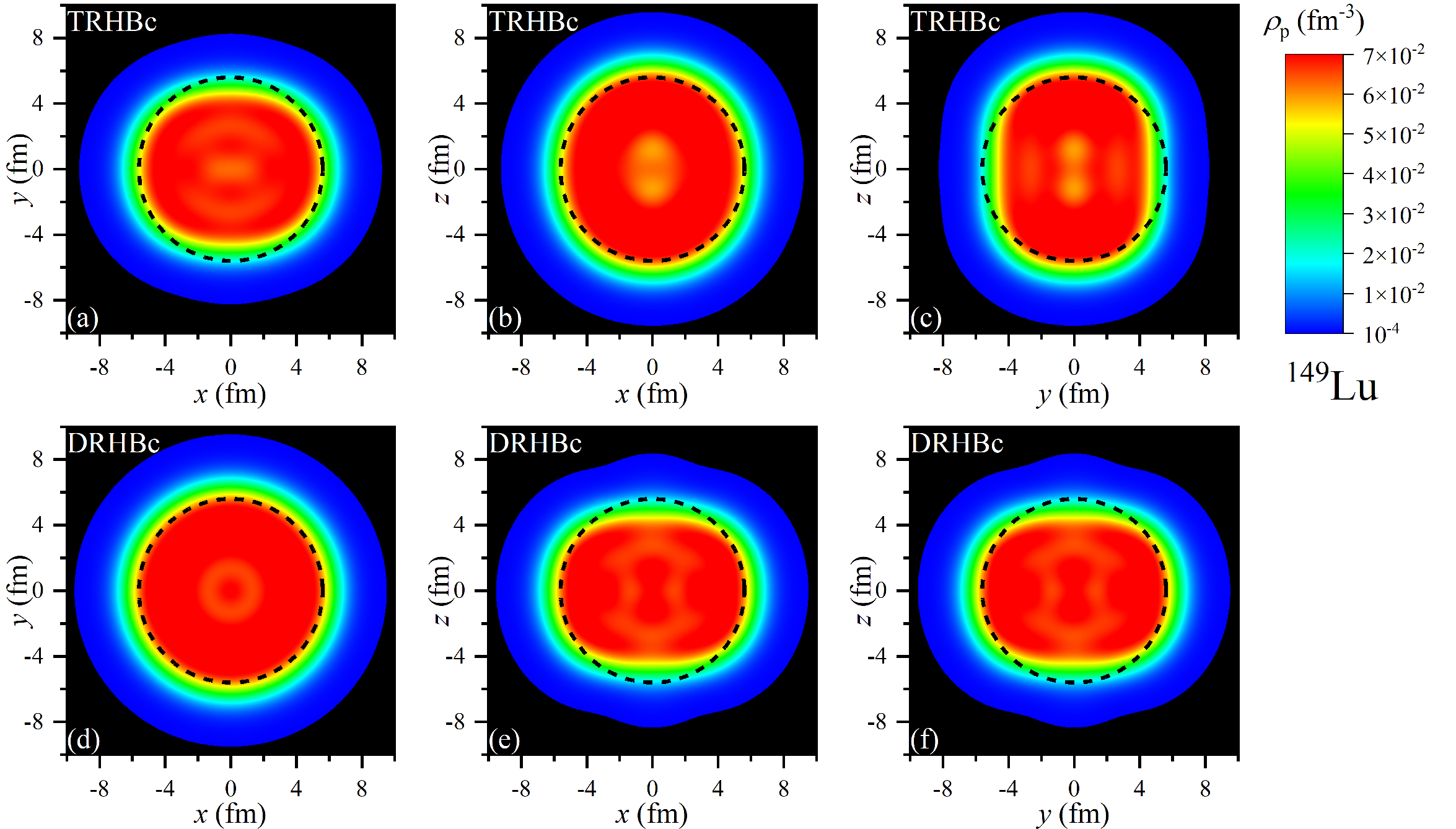}
  \caption{Proton density distributions of $^{149}$Lu in $xy$, $xz$, and $yz$ planes calculated from the TRHBc theory (upper panels) and the DRHBc theory (lower panels). In each plot, a circle in dotted line is drawn to guide the eye. 
  }\label{fig:rou}
\end{figure*}

With the consideration of triaxial deformation degrees of freedom, the TRHBc theory predicts that both $^{149}$Lu and its proton-emission daughter nucleus $^{148}$Yb are triaxially deformed.
The ground-state deformation parameters $(\beta,\gamma)$ are $(0.17, 31^\circ)$ and $(0.17, 35^\circ)$ for $^{149}$Lu and $^{148}$Yb, respectively.
Figure \ref{fig:potentialenergy} exhibits their potential energy surfaces (PESs), in which the absolute minima are indicated by red stars.
Due to the only one proton difference, the PES and the triaxial ground-state deformation for the daughter nucleus $^{148}$Yb are analogous to those for the parent nucleus $^{149}$Lu.
It can be clearly seen that the oblate deformations (red closed circles) predicted by the DRHBc theory are in fact saddle points in the PESs.
This demonstrates the importance of the inclusion of triaxial deformation in the microscopic description of $^{149}$Lu and $^{148}$Yb.
Although the energy difference between the saddle point and the absolute minimum is less than $0.4$ MeV in both PESs, the triaxial density distributions and potential barriers with $\gamma \gtrsim 30^\circ$ largely deviate from those with axial symmetry, which might affect the proton-emission half-life.

For an intuitive comparison, we display the proton density distributions of $^{149}$Lu in different planes by the TRHBc and DRHBc theories of Fig. \ref{fig:rou}.
With $z$ as the symmetry axis, the density distributions calculated by the DRHBc theory in the $xy$ plane shows rotational invariance, and those in $xz$ and $yz$ planes are identical.
Due to the predicted oblate shape, the density is more extended along $x$ and $y$ axes than that along $z$ axis, as shown in Figs. \ref{fig:rou}(e) and \ref{fig:rou}(f).
Namely, $z$ axis is the short one of the density distribution in the DRHBc theory.
However, this is not the case in the TRHBc theory that self-consistently incorporates triaxiality, {\it{i.e.}} the proton density distributions in three planes are different from each other.
It can be easily distinguished from Figs. \ref{fig:rou}(a) and \ref{fig:rou}(c) that $y$ is the short axis of the proton density distribution.
A detailed comparison between the density distribution in $xz$ plane and the dotted circle in Fig. \ref{fig:rou}(b) unveils that $z$ is the long axis while $x$ is the intermediate one.
Therefore, the spatial density distribution for $^{149}$Lu from the TRHBc theory is remarkably different from that from the DRHBc theory.
Definitely, the triaxially deformed potential barriers in the TRHBc theory also deviate from the axially symmetric ones in the DRHBc theory, which would affect the value of the decay width $\it{\Gamma}$.

\begin{table}[htpb]
\centering
\caption{Half-life of the proton emitter $^{149}$Lu. Note that for theories the orbital refers to the main component of the single-proton level occupied by the valence proton.}
\label{tb:result}
\resizebox{\linewidth}{!}{
\begin{tabular}{ccccc}
\hline
\hline
\multirow{2}{*}{$^{149}$Lu} &\multirow{2}{*}{Exp.~\cite{PhysRevLett.128.112501}} &   \multicolumn{3}{c}{NL3$^*$ (+ WKB)}  \\\cline{3-5}    
& & DRHBc & TRHBc &  \\
\hline
Orbital & $1h_{{11/2}}$ & $1h_{{11/2}}$ & $1h_{{11/2}}$\\
$\beta$ & / & $-0.18$  & 0.17\\
$\gamma$ & / & / & $31^\circ$\\
lg$\it{\Gamma}$ & / & $-14.86_{-0.09}^{+0.14}$ &  $-14.89_{-0.12}^{+0.12}$ \\
$S_F$ & / & 0.80 & 0.70 \\
$T_{1/2}$ (ns) & $450_{-100}^{+170}$ & $438_{-102}^{+136}$ & $504_{-116}^{+160}$\\
\hline\hline
\end{tabular}
}
\end{table}

With the nuclear structure inputs from the DRHBc and TRHBc theories, including the orbital angular momentum $l$ carried by the emitted proton and the spectroscopic factor $S_F$, the half-life of $^{149}$Lu can be estimated by the WKB approximation~\cite{XIAO2023138160}.
The structure properties and the calculated half-lives are listed in Table \ref{tb:result} in comparison with available data.
Following \cite{XIAO2023138160}, we employ the experimental $Q$ value and its error to compute the half-life and estimate the corresponding uncertainty.
In both theoretical results, the valence proton in $^{149}$Lu occupies a single-proton level with dominant $1h_{{11/2}}$ component, in agreement with the experimental data~\cite{PhysRevLett.128.112501}.
Furthermore, the estimated proton-emission half-lives are consistent with the measured data within uncertainties, which validates our calculations.
However, disregarding uncertainties, the predicted proton-emission half-life from the TRHBc + WKB model is longer than that from the DRHBc + WKB model by 66 ns. 
Surprisingly, it is found that the reduction of $S_F$ from $0.80$ to $0.70$ by including triaxial deformation contributes predominantly to the increase of $T_{1/2}$, while the decay widths calculated from the triaxially and axially deformed potential barriers are close to each other, which seems to contract with the speculation based on density distributions.

\begin{figure}[htpb]
\centering 
\includegraphics[width=0.48\textwidth]{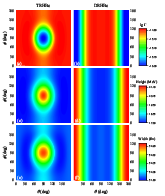}
\caption{The logarithm of the decay width, the height of potential barrier, and the width of potential barrier as functions of the orientation angle $\theta$ and the azimuthal angle $\phi$. The left and right panels exhibit results from the TRHBc + WKB and DRHBc + WKB models, respectively.}
\label{fig:GammaHeightWidth}
\end{figure}

To understand the obtained similar decay widths from the TRHBc + WKB and DRHBc + WKB models, we compare the calculated lg$\it\Gamma$ in different directions in the upper panels of Fig. \ref{fig:GammaHeightWidth}.
In the calculations of the TRHBc + WKB model shown in Fig. \ref{fig:GammaHeightWidth}(a), small decay widths appear in an ellipse-like region centered around $(\theta = 90^\circ, \phi=90^\circ)$.
As $\theta$ or $\phi$ deviates from the center, the decay width increases until lg$\it{\Gamma}$ reaches the maximum values of $\approx-14.5$.
The results of the DRHBc + WKB model in Fig. \ref{fig:GammaHeightWidth}(b) are only $\theta$-dependent due to the assumed axial symmetry and have completely different features. 
Large decay widths with lg$\it{\Gamma}$ $\gtrsim -14.5$ appear in the region of $45^\circ \lesssim \theta \lesssim 135^\circ$.
The decay width reduces with the decrease of $\theta$ smaller than $45^\circ$ or the increase of $\theta$ larger than $135^\circ$.
Note that the contribution of each $(\theta, \phi)$ point to the angle average is not equal because of a term $\sin\theta$ in the angular integral of Eq. \eqref{eq:finalWidth}.
While in the majority of the $\theta$-$\phi$ plane, the calculated decay widths from the TRHBc + WKB model are larger than those from the DRHBc + WKB model.
It occurs in an opposite way for $\theta$ close to $90^\circ$ with larger $\sin\theta$ values.
So it is the cancellation of these two effects that results in almost the same angle-averaged decay width with two models.

On the other hand, the significant discrepancy between the results shown in Figs. \ref{fig:GammaHeightWidth}(a) and \ref{fig:GammaHeightWidth}(b) indicates that triaxial deformation definitely affects the decay width in specific directions.
The heights and widths of the potential barrier are displayed in middle and lower panels of Fig. \ref{fig:GammaHeightWidth}, respectively.
The barrier height is defined as the peak value of the potential, and the barrier width refers to the distance between the second and third turning points of the potential with respect to the $Q$ value, $r_3-r_2$.
A negative correlation between the decay width and the barrier height/width is clearly revealed from both the TRHBc + WKB and DRHBc + WKB results.
This can be explained by the fact that, the higher or wider the barrier is, the more difficult it is for the valence proton to tunnel through.

\begin{figure*}[htpb]  
  \includegraphics[width=0.96\textwidth]{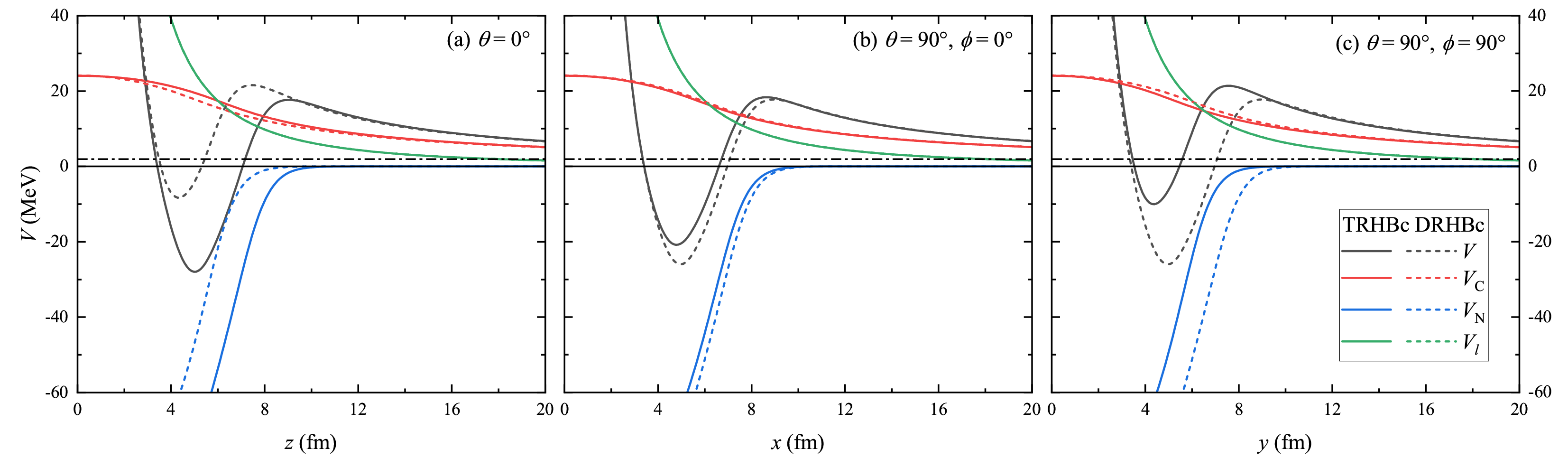}
  \caption{Potentials along $z$ ($\theta=0^\circ$), $x$ ($\theta=90^\circ$, $\phi=0^\circ$), and $y$ ($\theta=90^\circ$, $\phi=90^\circ$) axes. The solid and dotted lines show results from the TRHBc and DRHBc theories, respectively. The black, red, blue, and green lines represent the total, Coulomb, nuclear, and centrifugal potentials, respectively.
  }\label{fig:V}
\end{figure*}

To clarify the effects of triaxial deformation on the height and width of the potential barrier in $^{149}$Lu, we present the potential \eqref{eq:barrier} along $z$ ($\theta=0^\circ$), $x$ ($\theta=90^\circ, \phi=0^\circ$), and $y$ axes ($\theta=90^\circ, \phi=90^\circ$) in Fig. \ref{fig:V}.
The TRHBc and DRHBc results share the same centrifugal potential ($l = 5$ for $1h_{11/2}$).
Along each direction, there is no significant discrepancy between the Coulomb potentials for two theories.
As a result, the difference in barriers mainly stems from nuclear potentials.
It is then necessary to point out that since the Coulomb potential decays with $r$ gradually, the peak of the barrier is located where the nuclear potential approaches zero and its height decreases with this position away from the center.
A higher barrier also corresponds to a smaller $r_2$ and thus is wider, which explains the similarity of Figs. \ref{fig:GammaHeightWidth}(c, d) as Figs. \ref{fig:GammaHeightWidth}(d, f). 
In Fig. \ref{fig:V}(a) for $\theta = 0^\circ$, the nuclear potential diminishes to nearly zero at $z<8$ fm in the DRHBc theory, while it vanishes at $z \approx 10$ fm in the TRHBc theory.
This is due to a positive correlation between the nuclear density and potential. 
In the DRHBc calculations, $z$ is the short axis of the oblate density distribution, while it becomes the long axis of the triaxial density distribution in the TRHBc calculations.
Therefore, along this axis the barrier in the DRHBc theory is higher and wider than that in the TRHBc theory.
In Fig. \ref{fig:V}(c) for $\theta = 90^\circ$ and $\phi = 90^\circ$, the situation is just reversed as $y$ is the long (short) axis in the DRHBc (TRHBc) theory.
In Fig. \ref{fig:V}(b) for $\theta = 90^\circ$ and $\phi = 0^\circ$, the calculated barrier heights (widths) are close to each other.

\begin{figure}[htpb]  
  \includegraphics[width=0.48\textwidth]{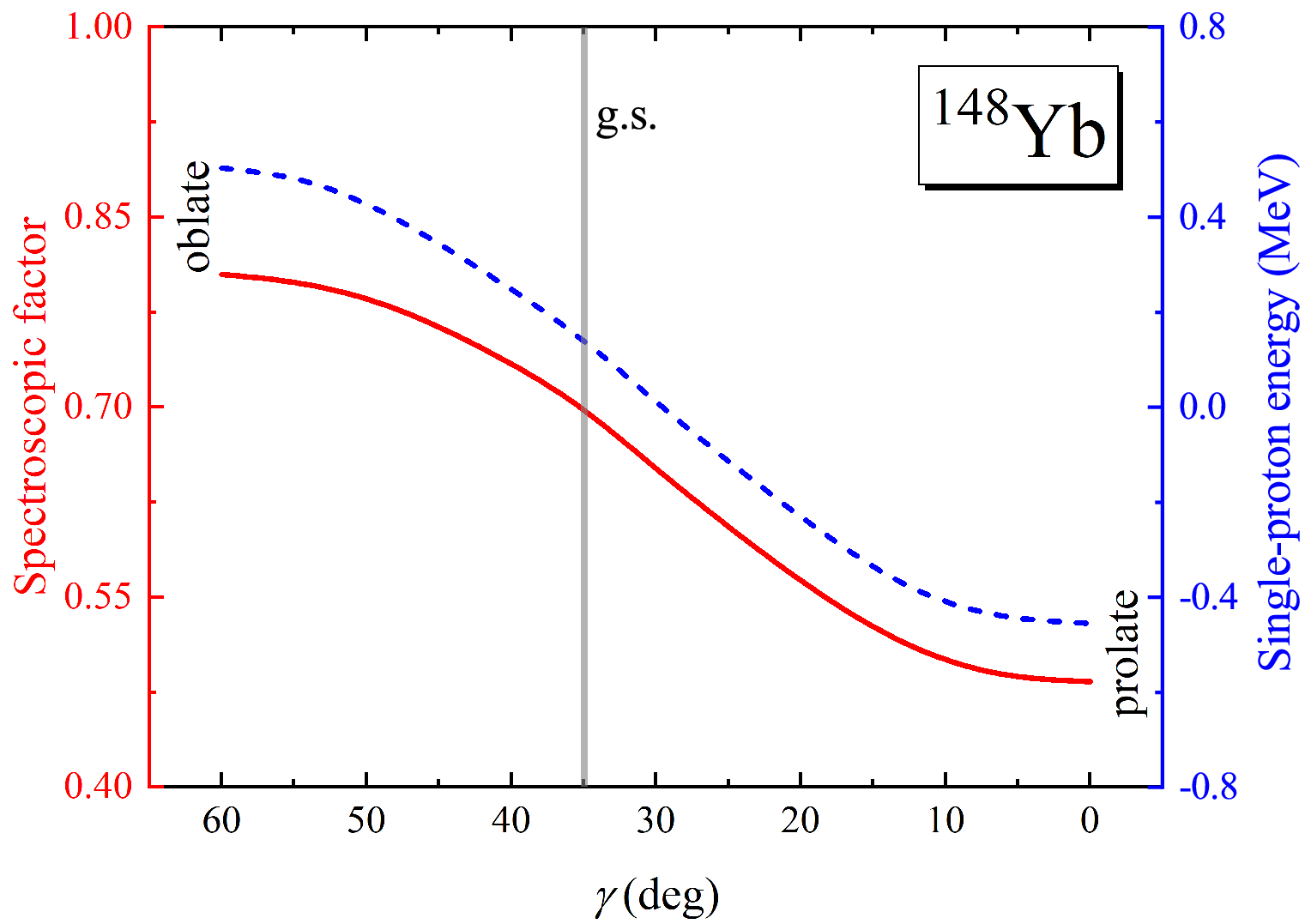}
  \caption{Spectroscopic factor (left vertical axis) and single-proton energy (right vertical axis) as functions of $\gamma$ for the daughter nucleus $^{148}$Yb. The vertical grey line corresponds to the ground-state (g.s.) triaxial deformation.
  }\label{fig:figure_SF}
\end{figure}

Finally, to explore the mechanism behind the reduction of the spectroscopic factor $S_F$ after considering triaxial deformation, we depict in Fig. \ref{fig:figure_SF} the evolution of the spectroscopic factor starting from the oblate side ($\gamma = 60^\circ$) to the prolate one ($\gamma = 0^\circ$).
A monotonic decline of $S_F$ can be seen from 0.8 to $\approx 0.48$ with the decrease of $\gamma$ from $60^\circ$ to $0^\circ$. 
The triaixal ground state in between corresponds to $S_F = 0.7$.
Given that $S_F$ is defined as the nonoccupation probability $u^2$ of the valence orbital after the proton emission, the evolution of its single-particle energy with triaxial deformation is also shown in Fig. \ref{fig:figure_SF} (right vertical axis).
Similar as $S_F$, the single-particle energy also demonstrates a monotonic decline with the decrease of $\gamma$.
Now, it can be understood, {\it{e.g.}}, in a simple BCS picture, that the orbital becomes more easily occupied with its decreasing single-particle energy.
Hence, with triaxial deformation included, not only the whole nuclear system but also the orbital involved in the proton emission become more bound, giving rise to the decrease of the spectroscopic factor.

\section{Summary}
\label{sec:summary}  

In conclusion, we revisit the recently identified proton emitter $^{149}$Lu utilizing the TRHBc + WKB approach and successfully reproduce the measured proton-emission half-life of $^{149}$Lu within experimental uncertainties.
Different from the previous studies indicating the coexistence of prolate and oblate shapes in $^{149}$Lu, triaxiality is taken into account self-consistently in the TRHBc theory.
In this scheme, the obtained PES unequivocally reveals a triaxial ground state characterized by ($\beta=0.17,\gamma=31^\circ$) for $^{149}$Lu.
Remarkably, the oblate ground-state deformation predicted by the DRHBc theory turns out a saddle point in the PES.
The triaxial ground-state density distribution significantly differs from the axially symmetric one.
In the former, $z$, $x$, and $y$ are respectively the long, intermediate, and short axes, whereas in the latter, $z$ is the short axis.
Consequently, pronounced discrepancies emerge in the triaxially and axially deformed nuclear potentials, which lead to distinct barrier heights and widths along specific directions.
Nonetheless, upon averaging contributions from all directions, we find that the angle-averaged decay width from the TRHBc theory approximately matches that from the DRHBc theory.
On the other hand, the inclusion of triaxiality results in enhanced binding of both the nuclear system and the proton-emitting orbital, yielding a reduced spectroscopic factor compared to that in the DRHBc calculations.
Ultimately, a slightly extended half-life for the proton emission of $^{149}$Lu is achieved when considering triaxial deformation degrees of freedom.

\section{Declaration of competing interest}

The authors declare that they have no known competing financial interests or personal relationships that could have appeared to influence the work reported in this paper.

\section{Data availability}

Data will be made available on reasonable request.

\section{Acknowledgement}

We thank Prof. Xiaohong Zhou for his useful discussions and careful reading of the manuscript.
This work was supported by the National Natural Science Foundation of China
(Grant Nos. 12175010 and 12305125) and Sichuan Science and Technology Program (Grant No. 2024NSFSC1356).

\bibliography{reference}

\end{document}